
\input phyzzx

\catcode`@=11
\def\space@ver#1{\let\@sf=\empty \ifmmode #1\else \ifhmode
   \edef\@sf{\spacefactor=\the\spacefactor}\unskip${}#1$\relax\fi\fi}
\def\attach#1{\space@ver{\strut^{\mkern 2mu #1} }\@sf\ }
\newtoks\foottokens
\newbox\leftpage \newdimen\fullhsize \newdimen\hstitle \newdimen\hsbody
\newif\ifreduce  \reducefalse
\def\almostshipout#1{\if L\lr \count2=1
      \global\setbox\leftpage=#1 \global\let\lr=R
  \else \count2=2
    \shipout\vbox{\special{dvitops: landscape}
      \hbox to\fullhsize{\box\leftpage\hfil#1}} \global\let\lr=L\fi}
\def\smallsize{\relax
\font\eightrm=cmr8 \font\eightbf=cmbx8 \font\eighti=cmmi8
\font\eightsy=cmsy8 \font\eightsl=cmsl8 \font\eightit=cmti8
\font\eightt=cmtt8
\def\eightpoint{\relax
\textfont0=\eightrm  \scriptfont0=\sixrm
\scriptscriptfont0=\sixrm
\def\rm{\fam0 \eightrm \f@ntkey=0}\relax
\textfont1=\eighti  \scriptfont1=\sixi
\scriptscriptfont1=\sixi
\def\oldstyle{\fam1 \eighti \f@ntkey=1}\relax
\textfont2=\eightsy  \scriptfont2=\sixsy
\scriptscriptfont2=\sixsy
\textfont3=\tenex  \scriptfont3=\tenex
\scriptscriptfont3=\tenex
\def\it{\fam\itfam \eightit \f@ntkey=4 }\textfont\itfam=\eightit
\def\sl{\fam\slfam \eightsl \f@ntkey=5 }\textfont\slfam=\eightsl
\def\bf{\fam\bffam \eightbf \f@ntkey=6 }\textfont\bffam=\eightbf
\scriptfont\bffam=\sixbf   \scriptscriptfont\bffam=\sixbf
\def\tt{\fam\ttfam \eightt \f@ntkey=7 }
\def\caps{\fam\cpfam \tencp \f@ntkey=8 }\textfont\cpfam=\tencp
\setbox\strutbox=\hbox{\vrule height 7.35pt depth 3.02pt width\z@}
\samef@nt}
\def\Eightpoint{\eightpoint \relax
  \ifsingl@\subspaces@t2:5;\else\subspaces@t3:5;\fi
  \ifdoubl@ \multiply\baselineskip by 5
            \divide\baselineskip by 4\fi }
\parindent=16.67pt
\itemsize=25pt
\thinmuskip=2.5mu
\medmuskip=3.33mu plus 1.67mu minus 3.33mu
\thickmuskip=4.17mu plus 4.17mu
\def\thinspace{\kern .13889em }
\def\negthinspace{\kern-.13889em }
\def\enspace{\kern.416667em }
\def\enskip{\hskip.416667em\relax}
\def\quad{\hskip.83333em\relax}
\def\qquad{\hskip1.66667em\relax}
\def\crr{\cropen{8.3333pt}}
\foottokens={\Eightpoint\singlespace}
\def\papersize{\SIZE\OFFSET\skip\footins=\bigskipamount}
\def\SIZE{\hsize=11.8truecm\vsize=17.5truecm}
\def\OFFSET{\voffset=-1.3truecm\hoffset=  .14truecm}
\message{STANDARD CERN-PREPRINT FORMAT}
\def\attach##1{\space@ver{\strut^{\mkern 1.6667mu ##1} }\@sf\ }
\def\PH@SR@V{\doubl@true\baselineskip=20.08pt plus .1667pt minus .0833pt
             \parskip = 2.5pt plus 1.6667pt minus .8333pt }
\def\author##1{\vskip\frontpageskip\titlestyle{\tencp ##1}\nobreak}
\def\address##1{\par\kern 4.16667pt\titlestyle{\tenpoint\it ##1}}
\def\andaddress{\par\kern 4.16667pt \centerline{\sl and} \address}
\def\abstract{\vskip2\frontpageskip\centerline{\tenrm Abstract}
              \vskip\headskip }
\def\cases##1{\left\{\,\vcenter{\Tenpoint\m@th
    \ialign{$####\hfil$&\quad####\hfil\crcr##1\crcr}}\right.}
\def\matrix##1{\,\vcenter{\Tenpoint\m@th
    \ialign{\hfil$####$\hfil&&\quad\hfil$####$\hfil\crcr
      \mathstrut\crcr\noalign{\kern-\baselineskip}
     ##1\crcr\mathstrut\crcr\noalign{\kern-\baselineskip}}}\,}
\Tenpoint
}
\def\Smallsize{\smallsize\reducetrue
\let\lr=L
\hstitle=8truein\hsbody=4.75truein\fullhsize=24.6truecm\hsize=\hsbody
\output={
  \almostshipout{\leftline{\vbox{\makeheadline
  \pagebody\makefootline}}}\advancepageno
     }
\special{dvitops: landscape}
\def\makeheadline{
\iffrontpage\line{\the\headline}
             \else\vskip .0truecm\line{\the\headline}\vskip .5truecm \fi}
\def\makefootline{\iffrontpage\vskip  0.truecm\line{\the\footline}
               \vskip -.15truecm\line{\the\date\hfil}
              \else\line{\the\footline}\fi}
\paperheadline={
\iffrontpage\hfil
               \else
               \tenrm\hss $-$\ \folio\ $-$\hss\fi    }
\paperstyle}
%
%
%
%
%
%
%
%
%
\newcount\referencecount     \referencecount=0
\newif\ifreferenceopen       \newwrite\referencewrite
\newtoks\rw@toks
\def\NPrefmark#1{\attach{\scriptscriptstyle [ #1 ] }}
\let\PRrefmark=\attach
\def\refmark#1{\relax\ifPhysRev\PRrefmark{#1}\else\NPrefmark{#1}\fi}
\def\refend{\refmark{\number\referencecount}}
\newcount\lastrefsbegincount \lastrefsbegincount=0
\def\refsend{\refmark{\count255=\referencecount
   \advance\count255 by-\lastrefsbegincount
   \ifcase\count255 \number\referencecount
   \or \number\lastrefsbegincount,\number\referencecount
   \else \number\lastrefsbegincount-\number\referencecount \fi}}
\def\refch@ck{\chardef\rw@write=\referencewrite
   \ifreferenceopen \else \referenceopentrue
   \immediate\openout\referencewrite=referenc.texauxil \fi}
%
{\catcode`\^^M=\active 
  \gdef\obeyendofline{\catcode`\^^M\active \let^^M\ }}%
%
{\catcode`\^^M=\active 
  \gdef\ignoreendofline{\catcode`\^^M=5}}
{\obeyendofline\gdef\rw@start#1{\def\t@st{#1} \ifx\t@st\blankend%
\endgroup \@sf \relax \else \ifx\t@st\bl@nkend \endgroup \@sf \relax%
\else \rw@begin#1
\backtotext
\fi \fi } }
{\obeyendofline\gdef\rw@begin#1
{\def\n@xt{#1}\rw@toks={#1}\relax%
\rw@next}}
\def\blankend{}
{\obeylines\gdef\bl@nkend{
}}
\newif\iffirstrefline  \firstreflinetrue
\def\rwr@teswitch{\ifx\n@xt\blankend \let\n@xt=\rw@begin %
 \else\iffirstrefline \global\firstreflinefalse%
\immediate\write\rw@write{\noexpand\obeyendofline \the\rw@toks}%
\let\n@xt=\rw@begin%
      \else\ifx\n@xt\rw@@d \def\n@xt{\immediate\write\rw@write{%
        \noexpand\ignoreendofline}\endgroup \@sf}%
             \else \immediate\write\rw@write{\the\rw@toks}%
             \let\n@xt=\rw@begin\fi\fi \fi}
\def\rw@next{\rwr@teswitch\n@xt}
\def\rw@@d{\backtotext} \let\rw@end=\relax
\let\backtotext=\relax

\newdimen\refindent     \refindent=30pt
\def\refitem#1{\par \hangafter=0 \hangindent=\refindent \Textindent{#1}}
\def\REFNUM#1{\space@ver{}\refch@ck \firstreflinetrue%
 \global\advance\referencecount by 1 \xdef#1{\the\referencecount}}
\def\refnum#1{\space@ver{}\refch@ck \firstreflinetrue%
 \global\advance\referencecount by 1 \xdef#1{\the\referencecount}\refend}

\def\REF#1{\REFNUM#1%
 \immediate\write\referencewrite{%
 \noexpand\refitem{#1.}}%
\begingroup\obeyendofline\rw@start}
\def\ref{\refnum\?%
 \immediate\write\referencewrite{\noexpand\refitem{\?.}}%
\begingroup\obeyendofline\rw@start}
\def\Ref#1{\refnum#1%
 \immediate\write\referencewrite{\noexpand\refitem{#1.}}%
\begingroup\obeyendofline\rw@start}
\def\REFS#1{\REFNUM#1\global\lastrefsbegincount=\referencecount
\immediate\write\referencewrite{\noexpand\refitem{#1.}}%
\begingroup\obeyendofline\rw@start}
\newif\ifmref  
\newif\iffref  
\def\xrefsend{\xrefmark{\count255=\referencecount
\advance\count255 by-\lastrefsbegincount
\ifcase\count255 \number\referencecount
\or \number\lastrefsbegincount,\number\referencecount
\else \number\lastrefsbegincount-\number\referencecount \fi}}
\def\xrefsdub{\xrefmark{\count255=\referencecount
\advance\count255 by-\lastrefsbegincount
\ifcase\count255 \number\referencecount
\or \number\lastrefsbegincount,\number\referencecount
\else \number\lastrefsbegincount,\number\referencecount \fi}}
\def\xREFNUM#1{\space@ver{}\refch@ck\firstreflinetrue%
\global\advance\referencecount by 1
\xdef#1{\xrefend}}
\def\xrefend{\xrefmark{\number\referencecount}}
\def\xrefmark#1{[{#1}]}
\def\xRef#1{\xREFNUM#1\immediate\write\referencewrite%
{\noexpand\refitem{#1 }}\begingroup\obeyendofline\rw@start}%
\def\xREFS#1{\xREFNUM#1\global\lastrefsbegincount=\referencecount%
\immediate\write\referencewrite{\noexpand\refitem{#1 }}%
\begingroup\obeyendofline\rw@start}
\def\rrr#1#2{\relax\ifmref{\iffref\xREFS#1{#2}%
\else\xRef#1{#2}\fi}\else\xRef#1{#2}\xrefend\fi}
\referencecount=0
\def\refbreak{\hfil\penalty200\hfilneg}
\def\paperstyle{\papers}
\paperstyle   
%
%
%
\def\slacpub{\afterassignment\slacp@b\toks@}
\def\slacp@b{\edef\n@xt{\Pubnum={NIKHEF--H/\the\toks@}}\n@xt}
\let\pubnum=\slacpub
\expandafter\ifx\csname eightrm\endcsname\relax
    \let\eightrm=\ninerm \let\eightbf=\ninebf \fi

\font\seventeencp=cmcsc10 scaled\magstep3

\newif\ifCONF \CONFfalse
\newif\ifBREAK \BREAKfalse
\newif\ifsectionskip \sectionskiptrue

%
%
%
%
\def\NuclPhysProc{
\let\lr=L
\hstitle=8truein\hsbody=4.75truein\fullhsize=21.5truecm\hsize=\hsbody
\hstitle=8truein\hsbody=4.75truein\fullhsize=20.7truecm\hsize=\hsbody
\output={
  \almostshipout{\leftline{\vbox{\makeheadline
  \pagebody\makefootline}}}\advancepageno
     }
\def\papersize{\SIZE\OFFSET\skip\footins=\bigskipamount}
\def\SIZE{\hsize=10.0truecm\vsize=27.0truecm}
\def\OFFSET{\voffset=-1.4truecm\hoffset=-2.40truecm}
\message{NUCLEAR PHYSICS PROCEEDINGS FORMAT}
\def\makeheadline{
\iffrontpage\line{\the\headline}
             \else\vskip .0truecm\line{\the\headline}\vskip .5truecm \fi}
\def\makefootline{\iffrontpage\vskip  0.truecm\line{\the\footline}
               \vskip -.15truecm\line{\the\date\hfil}
              \else\line{\the\footline}\fi}
\paperheadline={\hfil}
\paperstyle}
%
%
%
%

%
%
%
%
\def\Hopkins{\CONFtrue\paperheadline={\hfil}\reducefalse
\def\chapterfont{\bf}
\normalbaselineskip= 14.0pt plus .2pt minus .1pt
\def\papersize{\hsize=36.0pc\vsize=51pc\voffset 1.27truecm
    \hoffset -.09truecm\skip\footins=\bigskipamount}
\message{JOHNS HOPKINS FORMAT}
\normaldisplayskip= 18.0pt plus 5pt minus 10pt
\paperstyle }
%
%
%
%
\def\ReprintVolume{\smallsize
\def\papersize{\hsize=18.0truecm\vsize=23.1truecm\voffset -.73truecm
    \hoffset -.65truecm\skip\footins=\bigskipamount
    \normaldisplayskip= 20pt plus 5pt minus 10pt}
\message{REPRINT VOLUME FORMAT}
\paperstyle\baselineskip=.425truecm\parskip=0truecm
\def\makeheadline{
\iffrontpage\line{\the\headline}
             \else\vskip .0truecm\line{\the\headline}\vskip .5truecm \fi}
\def\makefootline{\iffrontpage\vskip  0.truecm\line{\the\footline}
               \vskip -.15truecm\line{\the\date\hfil}
              \else\line{\the\footline}\fi}
\paperheadline={
\iffrontpage\hfil
               \else
               \tenrm\hss $-$\ \folio\ $-$\hss\fi    }
\def\sectionfont{\bf}    }
%
%
%
%
\def\SIZE{\hsize=15.73truecm\vsize=23.11truecm}
\def\OFFSET{\voffset=0.0truecm\hoffset=0.truecm}
\message{DEFAULT FORMAT}
\def\papersize{\SIZE\OFFSET\skip\footins=\bigskipamount
\normaldisplayskip= 35pt plus 3pt minus 7pt}
\Pubnum={\rm NIKHEF--H/\the\pubnum }
\def\title#1{\vskip\frontpageskip\vskip .50truein
     \titlestyle{\seventeencp #1} \vskip\headskip\vskip\frontpageskip
     \vskip .2truein}
\def\author#1{\vskip .27truein\titlestyle{#1}\nobreak}

\def\p@bblock{\begingroup \tabskip=\hsize minus \hsize
   \baselineskip=1.5\ht\strutbox \topspace+2\baselineskip
   \halign to\hsize{\strut ##\hfil\tabskip=0pt\crcr
  \the \Pubnum\cr
}\endgroup}
\def\makefootline{\iffrontpage\vskip .27truein\line{\the\footline}
                 \vskip -.1truein
              \else\line{\the\footline}\fi}
\paperfootline={\iffrontpage\message{FOOTLINE}
\hfil\else\hfil\fi}

\def\abstract{\vskip2\frontpageskip\centerline{\twelvebf Abstract}
              \vskip\headskip }

\paperheadline={
\iffrontpage\hfil
               \else
               \twelverm\hss $-$\ \folio\ $-$\hss\fi}
%
%
\def\nup#1({\refbreak\ Nucl.\ Phys.\ $\underline {B#1}$\ (}
\def\plt#1({\refbreak\ Phys.\ Lett.\ $\underline  {#1}$\ (}
\def\cmp#1({\refbreak\ Commun.\ Math.\ Phys.\ $\underline  {#1}$\ (}
\def\prp#1({\refbreak\ Physics\ Reports\ $\underline  {#1}$\ (}
\def\prl#1({\refbreak\ Phys.\ Rev.\ Lett.\ $\underline  {#1}$\ (}
\def\prv#1({\refbreak\ Phys.\ Rev. $\underline  {D#1}$\ (}
\def\und#1({            $\underline  {#1}$\ (}
%
%

\def\rB{\hfil\penalty1000\hfilneg}
%
%
\hyphenation{sym-met-ric anti-sym-me-tric re-pa-ra-me-tri-za-tion
Lo-rentz-ian a-no-ma-ly di-men-sio-nal two-di-men-sio-nal}
%
%
%
%

%


%

%
\def\boxit#1{\vbox{\hrule\hbox{\vrule\kern3pt
\vbox{\kern3pt#1\kern3pt}\kern3pt\vrule}\hrule}}
\message{ by V.K, W.L and A.S}
\catcode`@=12
\paperstyle


\def\ScYg{\rrr\ScYg{A.N.~Schellekens and S.~Yankielowicz,
Int.~J.~Mod.~ Phys.~\und{A5} (1990) 2903.}}
\def\EVeA{\rrr\EVeA{E.~Verlinde,
\nup300 (1988) 360.}}
\def\ScYA{\rrr\ScYA{
A.N.~Schellekens and S.~Yankielowicz,
\nup 327 (1989) 673; \rB
\plt B227 (1989) 387.}}
\def\CIZ {\rrr\CIZ {
A.~Cappelli, C.~Itzykson     and J.-B.~Zuber,
\nup280 (1987)  445;\rB \cmp113 (1987) 1.}}
\def\Fuch{\rrr\Fuch{J.~Fuchs,
\cmp136 (1991) 345.}}
\def\FGSS{\rrr\FGSS {J. Fuchs, B. Gato-Rivera, A.N. Schellekens and C.
Schweigert,
\plt B334 (1994) 113.}}
\def\dBG{\rrr\dBG{J. de Boer and J. Goeree, \cmp 139 (1991) 267.}}
\def\CoGa{\rrr\CoGa{A. Coste and T. Gannon, \plt B323 (1994) 316.}}
\def\Gan{\rrr\Gan{T. Gannon,
\cmp{161}(1994){233}.}}
\def\GanB{\rrr\GanB{T. Gannon, {\it The Classification
of $SU(N)_k$ automorphism invariants}, hep-th 9212060.}}
\def\GanA{\rrr\GanA{T. Gannon, \nup 396 (1993) 708.}}
\def\FSch{\rrr\FSch {J.~Fuchs and C.~Schweigert, Ann. Phys. 234 (1994) 102.}}
\def\FSS{\rrr\FSS{ J. Fuchs, A.N. Schellekens and C. Schweigert, preprint
NIKHEF-94/31.}}
\def\FSSb{\rrr\FSSb{J. Fuchs, A.N. Schellekens and C. Schweigert, NIKHEF
preprint, in preparation.}}
\def\Aut{\rrr\Aut{D.~Bernard, \nup 288 (1987) 628;\rB D.~Altschuler, J.~Lacki
and P.~Zaugg, \plt B205 (1988) 281;\rB G.~Felder, K.~Gawedzki and A.~Kupiainen,
\cmp 117 (1988) 127.}}
\def\Ehol{\rrr\Ehol{W.~Eholzer, Int. J. Mod. Phys. 8 (1993) {3495}.}}

\Hopkins
\def\Zbf{{\bf Z}}
\def\Q{{\cal Q}}
\def\papersize{\hsize=39.0pc\vsize=55pc\voffset -0.10truecm
    \hoffset -.09truecm\skip\footins=\bigskipamount}
\paperstyle
\CONFfalse
\ifCONF
\message{CONFERENCE VERSION}
\line{\hfill}
\vskip 1.truecm
\line{\hfil \fourteenbf Modular Invariance and (Quasi)-Galois symmetry\hfil}
\line{\hfil \fourteenbf in
 Conformal
Field Theory\hfil}
\vskip 1.truecm
\line{\hfil \tenrm A.N. SCHELLEKENS\hfil}
\vskip .2truecm
\line{\hfil \tenpoint\it NIKHEF-H, P.O. Box 41882, 1009 DB Amsterdam,
The Netherlands  \hfil}
\vskip .5truecm
\line{\hfil \tenrm ABSTRACT \hfil}
\vskip .2truecm
\input tables
\thicksize=0pt \thinsize=0pt
\tablewidth=1.truecm
\parasize=30pc
\begintable
\para{\tenpoint A brief heuristic explanation is given of recent work [1,2]
with J\"urgen Fuchs, Beatriz Gato-Rivera and Christoph Schweigert
on the
construction of modular invariant partition functions from Galois
symmetry in conformal field theory. A generalization, which we call
quasi-Galois symmetry [3], is also described. As an application of the
latter, the invariants of the exceptional algebras at level $g$
(for example $E_8$ level 30) expected
from conformal embeddings are presented.
}\cr
\endtable
\singlespace
\baselineskip=15pt
\else
\def\papersize{\hsize=36.0pc\vsize=55pc\voffset -0.70truecm
    \hoffset .65truecm\skip\footins=\bigskipamount}
\paperstyle

\pubnum={94-38}
\date{November 1994}
\pubtype{CRAP}
\titlepage
\message{TITLE}
\title{\fourteenbf Modular Invariance and (Quasi)-Galois symmetry \break in
Conformal
Field Theory\foot{\ Presented at the
 International Symposium on the
Theory of Elementary Particles
Wendisch-Rietz, August 30 - September 3, 1994}}

\author{A. N. Schellekens}
\vskip 0.3truein
\vskip .2truecm
\line{\hfil \tenpoint\it NIKHEF-H, P.O. Box 41882, 1009 DB Amsterdam,
The Netherlands  \hfil}
\vskip 0.6truein
\abstract
  A brief heuristic explanation is given of recent work [1,2]
with J\"urgen Fuchs, Beatriz Gato-Rivera and Christoph Schweigert
on the
construction of modular invariant partition functions from Galois
symmetry in conformal field theory. A generalization, which we call
quasi-Galois symmetry [3], is also described. As an application of the
latter, the invariants of the exceptional algebras at level $g$
(for example $E_8$ level 30) expected
from conformal embeddings are presented.

\endpage
\paperheadline={
\iffrontpage\hfil
               \else
         \twelverm\hss $-$\ \folio\ $-$\hss\fi    } \paperstyle
\pagenumber=1
\fi

\ifCONF
 \def\chapter#1{\par
 \penalty-300 \vskip\chapterskip
   \spacecheck\chapterminspace
   \chapterreset
    \titlestyle{\hskip -.5truecm\twelvebf\chapterlabel.~#1\hfill\hfill\hfill}
   \nobreak\vskip\headskip \penalty 30000
   }
\fi

\chapter{Introduction\hfill}

\referencecount=3

Since the title of this conference mentions ``elementary particle physics",
I will begin by sketching the relation between that field and the subject
of this talk. The usual argument goes like this: ``Elementary particles
and their interactions are described by string theory. String theory may
be unique, but it has an extremely large number of vacua, and at most one
of those vacua corresponds to the standard model. Each vacuum
corresponds to a two-dimensional conformal field theory. In addition to
conformal invariance, such a theory may have additional invariances, and
it may happen that the symmetry is so powerful that the entire content of
the theory can be organized in terms of a finite number of representations.
Then the conformal field theory is called rational. Knowing all rational
conformal field theories may give sufficient information about the full
set of conformal field theories, either because the rational conformal field
theories cover the parameter space densely, or at least because one can
reach all theories by making small  perturbations around the
rational ones. All rational conformal field theories are cosets $G/H$ (where
$G$ is a WZW-model and $H$ a gauged subgroup) or orbifolds of cosets.
The problem of classifying all string vacua begins thus with the
classification of all ungauged WZW-models."

Many of the statements in the foregoing line of arguments are highly
questionable. We don't know whether string theory, or our present way
of thinking about it, has anything to do with nature, nor whether
RCFT can give us sufficient information about all its vacua. The
classification of RCFT's is still an unsolved problem, and the conjecture
that they are all somehow related to coset theories at least
requires some amendments. In addition, in many cases we do not even know
how to work out the characters, the partition functions, the fusion
rules and even the spectrum of coset theories because of fixed points
in the field identification. Furthermore solving the classification problem
for WZW models does not solve the classification problem for cosets.
Finally, the classification problem of WZW itself is probably
unsolvable. It may well be solved during our lifetime for the simple
WZW models, but this does not imply a classification for semi-simple
models. The classification problem  does not factorize, and
entirely new solutions can appear for tensor products of simple theories
that cannot be anticipated, as far as we know.

Nevertheless, the classification problem of WZW-models has
intrigued several people. Before asking why, let us formulate the
problem. Algebraically, a WZW-model is described in terms of a
combination of the Virasoro algebra of conformal transformations,
generated by currents of spin 2, and a Kac-Moody algebra
generated by spin 1 currents. The generators $T^a_n$
of a Kac-Moody algebra satisfy $$ [ T^a_n, T^b_m ] = i f^{abc} T^c + \half k m
\delta_{n,-m}\delta^{ab}  \ ,$$
where $f^{abc}$ are the structure constants of a  Lie-algebra, and
$k$ an integer  (in fact there such an integer for every simple factor of the
Lie-algebra.)
All (untwisted, simple) KM-algebras are thus characterized by
a type $\rm T=A,B,C,D,E,F$ or $\rm G$, a rank $r$ and a level $k$, and
they will be denoted ${\rm T}_{r,k}$.

There is some additional freedom in putting the conformal
field theory together. At one loop level, the differences manifest
themselves in the partition function. One-loop partition functions can
be written on the one hand as path-integral of the conformal field
theory on a torus, and on the other hand as a trace over the
exponentiated Hamiltonian. The latter expression has the form
$$ P(\tau,\bar\tau) = \Tr e^{2\pi i \tau H_L} e^{-2\pi i \bar \tau H_R} \ ,$$
where $H_L$ and $H_R$ are the Hamiltonians
(the zero-mode generators of the conformal transformations, up to a constant)
of the left- and right-moving
modes on the two-dimensional surface. Their
eigenvalues are called conformal weights.

The parameter $\tau$ describes the inequivalent shapes of the torus.
Equivalences due to reparametrizations or rescalings have been
removed from the path integral by gauge-fixing.
In string theory one must integrate over this parameter, but it turns out
that not all values of $\tau$ are really inequivalent. There are still some
global reparametrizations that have not been removed yet. This divides the
complex upper half plane in which $\tau$ lives in an infinite number of
equivalent regions, and if one integrates over one such region all tori
have been properly included. However, this procedure is only correct if
the integral does not depend on the region.

The different regions are mapped into each other by a discrete group
called the modular group, which is isomorphic to $SL(2,\Zbf)$. This
group is generated by two transformations
$$\eqalign{ S:   \tau &\rightarrow -{1\over\tau} \cr
T:   \tau &\rightarrow {\tau+1} \cr} $$
The requirement of modular invariance is now that $P(\tau,\bar\tau)$ is
invariant under these two transformations, and hence under the full
group. [The requirement that not just the integral, but the integrand
itself should be invariant arises since we should not just consider
the one-loop vacuum amplitude, but all correlators at any loop order].

Using the symmetries of  WZW we can
combine all physical states into a finite number of representations, each
of which is infinite-dimensional. Then the partition function takes the
following form
$$ P(\tau,\bar\tau) = \sum_{i,j} M_{ij} {\cal X}_i (\tau){\cal X}_j^*(\bar\tau)
\ , $$
where $M_{ij}$ is a set of non-negative integer multiplicities for
the left representation $i$ with the right representation $j$, and
${\cal X}_i$ is the (Virasoro) character of the representation $i$. These
characters transform in the following way under modular transformations
$$ \eqalign{ S:   {\cal X}_i &\rightarrow S_{ij} {\cal X}_j\cr
 T:   {\cal X}_i &\rightarrow T_{ij} {\cal X}_j\ , \cr} $$
where $T$ is a diagonal unitary matrix and $S$ a symmetric unitary matrix.
The condition for modular invariance is now simply
$$ [S,M] = [ T,M] = 0 \ ,\eqn\Cond$$
combined with the condition that $M$ be integer and non-negative.
Furthermore to have a unique ground state (denoted by ``0")
one requires
 that $M_{00}=1$.

At higher loops
some additional requirements have to be satisfied, but this is not expected
to affect the result very strongly. Nevertheless one should keep in mind
that solving the purely algebraic conditions \Cond\ does not yet
guarantee the existence of a corresponding conformal field theory.

A trivial solution to \Cond\ is $M={\bf 1}$, which is called the diagonal
invariant.
Several years ago, Cappelli, Itzykson and Zuber \CIZ\ found the general
solution for the Lie-algebra $A_1$ at arbitrary level $k$. It might have
seemed reasonable to expect the general solution to be known within
a year or so, but this has not happened. Just two years ago Gannon \Gan\
completed the classification for $A_2$ and arbitrary $k$, but although
some further progress was made, the complete classification
problem is still far from solved, even for simple algebras.

Why bother? As explained earlier, it would be ridiculous to claim that
solving this problem would produce a tremendous breakthrough in
understanding the vacuum structure in string theory. Nor would it be
of overwhelming importance for statistical mechanics, where the same
problems were formulated, mainly due to the work of Cardy. The
attraction of the problem is probably somewhat similar to that of
Fermat's conjecture: very simple to formulate, yet very hard to solve. This
is not to suggest that the problem of classifying all simple WZW modular
invariants is as important, as deep, or as difficult as Fermat's problem, only
that the motivation is somewhat similar. Just as in that case, the problem
itself may be far less important than the new insights one gains and the
methods one develops while trying to solve it.

There is a good example of that already.
A very general class of non-diagonal modular invariants is the one
generated by simple currents \ScYA.
Simple currents were
discovered by studying WZW modular invariants. In the special case
of WZW models almost all simple currents correspond to
automorphisms of the extended Dynkin diagrams, and the
partition functions can be viewed in terms of strings on non-simply
connected group manifolds, obtained by identifying points on the
universal covering group related by elements of the center of the
group \Aut.
But the notion of simple currents abstracted from
this is more general, and already WZW models
themselves provide an example, namely the simple current of $E_{8,2}$, which
has no known interpretation in terms of global properties of the group
manifold (among WZW-models this is the only exception \Fuch, but one
exception is sufficient to prove the point).

Galois symmetry could have been discovered in a similar way, although
historically this is not what happened. Our own involvement
\FGSS\
with this
subject started with an observation about the fusion rules. The
fusion rules of a CFT give the number of allowed three-point
couplings between certain representations. They have the form
$$ R_i \times R_j = \sum_k N_{ij}^k R_k\ , $$
where $R_i$ is a representation (or a ``primary field")
and $N_{ij}^k$ a non-negative integer.

It was pointed out by E. Verlinde \EVeA\ that $N_{ij}^k$ can be written in
terms
of the matrix $S$:
$$ N_{ij}^k = \sum_n { S_{in} S_{jn} S_{kn}^* \over S_{0n} }$$
In this formula the identity representation 0 plays a special r\^ole.
If the conformal field theory is unitary there is a natural choice for
this representation, namely the one with smallest conformal weight.
However in work on fixed points of simple
currents \ScYg\ we were confronted with $S$-matrices that had all the right
properties to be a modular group representation, but did not
correspond to a unitary CFT.
I will not attempt to explain fixed point resolution  here, but the main point
is that one needs a set of matrices $S_{\rm fix}$ ad $T_{\rm fix}$ that
form a representation of the modular group.
In most cases these
fixed point resolution matrices are themselves ``$S$-matrices''
of WZW-models, but there are exceptional cases where an unknown matrix
had to be used. These matrices are labelled by two positive integers
$n$ and $m$, and could be identified for $n=1$ and $m=1$ with the
matrices $S$ of  non-unitary minimal Virasoro theories. The matrix for
$n=m=2$ was constructed numerically\foot{meanwhile
it is known how to construct these
matrices for arbitrary $n$ and $m$ \FSch.}in \ScYg,  and
could not be identified with any   known CFT.

To  identify a conformal field theory that might correspond to this
matrix we tried to determine which of the six fields was the identity.
These
``CFT"'s, which still have not been identified, were tentatively called
${\cal B}_{m,n}$.
Our hope was to use the fusion rule formula to determine which field to
use. A priori one would expect that using the wrong field "0" would
lead to non-integer, fractional or  infinite fusion rule
coefficients.

Surprisingly, in the example we considered any choice seemed to be
equally good. In all cases the fusion rule coefficients turned out
to be integer, though in no cases were they positive. However, it was
always possible to find a set of signs $\epsilon(i)$ and a new
matrix $S'_{ij}=\epsilon(i) \epsilon(j) S_{ij}$ that made all the
coefficients positive. There was a second surprise: the fusion rules
turned out to be identical up to some permutation of all the fields.

Inspired by this observation we
investigated a few WZW models in a similar way, and we found that
some of the fields could play the r\^ole of the identity in the
above sense. This was not pursued further at the time until
recently, after the construction of the matrices $S$ for ${\cal B}_{m,n}$
for
arbitrary $n$ and $m$ (as far as we can tell now
these have indeed such a symmetry
under a certain signed cyclic permutation of
all fields).  Meanwhile the symmetry was discovered independently
by Eholzer \Ehol.

It is natural to look now for some underlying symmetry of $S$, and
it turns out that indeed such a symmetry exists. Consider a
matrix
$\Pi$ of the form
$$ \Pi_{ij} = \epsilon(i) \delta_{i,\pi(j)}\  = \epsilon(\pi^{-1}(j))
\delta_{\pi^{-1}i,j} \  , $$
where $\pi$ is a permutation and $\epsilon(i)$ a set of signs.
Suppose the matrix $S$ satisfies
$$ S = \Pi S \Pi\ , \eqn\SRel$$
which implies that some elements of $S$ are pairwise equal, up to
signs. If one inserts this into the Verlinde formula, one finds
$$ N_{ij}^k = \epsilon(i)\epsilon(j)\epsilon(k)\epsilon(0)\
N_{\pi(i)\pi(j)}^{\pi(k)}[\pi(0)]\ , \eqn\FusionRel$$
where the argument between square brackets indicates the choice of vacuum.
This is exactly the behavior observed above.
The signs can indeed be
removed by redefining $S$ in the way discussed above ($\epsilon(0)$ can be
set equal to 1 by an overall sign choice).

One can easily imagine other symmetries of $S$ that would lead to
\FusionRel,
for example one could replace one of the two matrices $\Pi$ in \SRel\ by
any other signed permutation $\Pi'$,
but empirically \SRel\ is
 the symmetry
responsible for the observed relation \FusionRel. This relation among
the element of $S$ can also be written as
$$  \Pi^T S =  S \Pi  \ ,\ \hbox{and} \ \ \ \ \ \Pi S = S \Pi^T \eqn\PiTPi $$
because $\Pi$ is an orthogonal matrix. This implies that the matrix
$\Pi + \Pi^T$ commutes with $S$. This is a useful step towards finding
modular invariant partition functions, although unfortunately commutation
with $T$ and positivity are not automatically guaranteed.

In general $S$ may have many symmetries of the type \SRel, but a
very large and interesting subclass is obtained from Galois symmetry.
In principle Galois symmetry of conformal field
theory might have been discovered in this way, but actually it had already
been found several years earlier by de Boer and Goeree \dBG. The basic
observation starts with writing the fusion rules in the following way
   $$ {S_{i\ell} \over S_{0\ell}}\, {S_{j\ell} \over S_{0\ell}} =
  \sum_k {\cal N}_{ij}^{\ k}\, {S_{k\ell} \over S_{0\ell}}  \eqn\odim $$
This means that the {\it generalized quantum dimensions}
 ${S_{j\ell} \over S_{0\ell}}$ furnish one-dimensional representations
of the fusion algebra. Furthermore they exhaust the set of
one-dimensional representations. Now suppose we have a map
$\sigma$
that takes complex numbers to complex numbers, that leaves
integers fixed, and that respects sums and products, $\ie\
\sigma(ab)=\sigma(a)\sigma(b)$ and $\sigma(a+b)=\sigma(a)+\sigma(b)$.
Applying such a map to both sides of $\odim$ we find then that
$\sigma({S_{j\ell} \over S_{0\ell}})$ is a  one-dimensional representation
of the fusion algebra. Hence it must be one of the representations
we already know, \ie\ $\sigma$ induces a permutation
$\dot{\sigma}$ of the labels $\ell$:
 $$\sigma({S_{j\ell} \over S_{0\ell}}) = {S_{j\dot{\sigma}(\ell)} \over
 S_{0\dot{\sigma}(\ell)}}\eqn\GaloisMap$$

An example of a map with such properties is complex conjugation, but
there is a more general class one can consider. These are the
Galois transformations. Complex conjugation means interchanging the
roots of the polynomial $x^2 + 1 $ = 0. This polynomial has rational
coefficients, but does not have roots within the rational numbers. One
has to extend the field $\Q$ to $\Q+i\Q$ to solve the polynomial equation.
This new field has a symmetry $a+ib \rightarrow a-ib$. This $\Zbf_2$ symmetry
is called the Galois group of the field extension. This idea can be
generalized to any polynomial with rational coefficients. For any such
polynomial there is a minimal extension $L$ of $\Q$  in which the
polynomial has roots. For such an extension there exists a discrete
group ${\cal G}(L/\Q)$ of automorphisms of $L$ that fix $Q$, called the Galois
group. The elements of this group leave rational numbers invariant and
act as permutations (possible trivial ones) on the roots of the polynomial.
Furthermore they respect the all defining properties of the field, in
particular
addition and multiplication. These symmetries can then be applied to
obtain relations of the form \GaloisMap.

In order to use Galois transformations it is important to know that the
generalized quantum dimensions are indeed roots of some polynomial
with rational coefficients. This is true because from $\odim$ one
sees that they are eigenvalues of the  integer matrices $(N^i)_{jk}=N_{ij}^k$.
Hence they are roots of the characteristic polynomials, and because
the matrices are integer
these polynomials do indeed have rational coefficients.

The next step was made by Coste and Gannon \CoGa. They observed that
the matrix elements of $S$ themselves belonged to a slightly larger
extension of the rational numbers, containing $L$. Furthermore they
showed that one can define a transformation similar to
\GaloisMap\ for the matrix elements of $S$:
$$ \sigma  S_{i,\ell} = \epsilon_{\sigma}(\ell) S_{i,\dot{\sigma}(\ell)} $$
Since $S$ is symmetric it is now easy to see that the Galois group
must be abelian\foot{This had already been proved for ${\cal G}(L/\Q)$ by
de Boer and Goeree, but the proof is slightly more complicated
in that case.}(two distinct elements can each act on a different index of
$S$), and furthermore one sees immediately that
$$ S_{ij} = \sigma\sigma^{-1} S_{ij}=
\epsilon_{\sigma}(i)\epsilon_{\sigma^{-1}}(j)
S_{\dot{\sigma}(i)\dot{\sigma}^{-1}(j)}\ . $$
This is precisely \SRel.

The fact that the Galois group is abelian has an important consequence
due to a theorem of Kronecker and Weber: it implies that the
matrix elements of $S$ are elements of a cyclotomic field. This is
an extension of the rational numbers by the powers of a certain
root of unity. Suppose the field is generated by $t=e^{2\pi i \over N}$.
Then Galois transformations correspond to sending
$t \rightarrow t^l$ for some scale factor $l$. For this to be an automorphism
of the field one has to require that $l$ and $N$ are relatively prime.

Up to now everything was valid for arbitrary conformal field theories.
Let us now be more specific and discuss WZW-models. Then a
formula exists for the matrix $S$ :
$$ S_{\lambda\mu}= {\cal N} \sum_w \epsilon(w) e^{2\pi i {w(\lambda+\rho)\cdot
(\mu+\rho) \over k+g }} \ ,\eqn\Kap $$
where $g$ is the dual Coxeter number and $\rho$ the Weyl vector,
whose Dynkin labels are $(1,1,1,1,\ldots,1,1)$; ${\cal N}$ is a normalization.
The sum is over the "horizontal" Weyl group (the Weyl group of the
Lie algebra.)
If ${\cal N}$ were rational the order $N$ of the cyclotomic field would
be determined completely by the denominator of the exponent. In general
${\cal N}$ is not rational, but this only leads to a further
extension of the cyclotomic field that is not relevant for our purposes.
The labels $\lambda$ and $\mu$  in \Kap\ are the highest weights of two
representations.
The  order  $N$ of the cyclotomic field (as defined above) is thus $M(k+g)$,
where $M$ is
the common denominator of the inner products among weights.
Galois transformations correspond thus to those rescalings $l$ of
the exponent so that  $l$ and $M(k+g)$ are relatively prime.

A familiar result for ordinary simple Lie algebras is that any weight can be
rotated into any preferred Weyl chamber by some suitable Weyl rotation.
This is not different for Kac-Moody  (affine) algebras. Only in this case there
is
an extra restriction on the Weyl chambers, which depends on $k$.
Due to this restriction the affine Weyl chambers are finite in size. This is
illustrated in fig.1 (next page), where the big triangle shows one affine Weyl
chamber
for $A_2$ level 5.
The Kac-Moody highest weights at level 5 are precisely
the weights in this triangle,
including the boundary.
It is again true that any weight can be transformed back
into the affine Weyl chamber, but now one has to use the affine
Weyl group, which contains in addition to the normal Weyl group also
refections with respect to the dashed line of the correct level (these lines
are indicated in the figure).

If we were to ignore the shift by $\rho$ in \Kap\
we could transform the scaled
weight
$l\mu$ back to the affine Weyl chamber by means of affine
Weyl transformations at level $k$. It is easy to see
that horizontal Weyl transformations can be removed by shifting the sum over
the horizontal Weyl group. This may change the sign of the overall factor
$\epsilon(w)$
by a sign that depends on $\mu$.  However, the
additional transformations
in the affine Weyl group change the exponent unless they are generated
by reflections with respect to the dashed line of level $k+g$.

The shift by $\rho$ implies that we are not scaling $\mu$ but $\mu+\rho$.
In the figure this is illustrated for $k=2$:
the shifted affine Weyl chamber for $k=2$ is the dark triangle, and
it is shown embedded in the unshifted $k=5$ Weyl chamber (large triangle).

\vskip .7truecm
\let\picnaturalsize=N
\def\picsize{4.8in}
\def\picfilename{A2.eps}
\ifx\nopictures Y\else{\ifx\epsfloaded Y\else\input epsf \fi
\let\epsfloaded=Y
\centerline{\ifx\picnaturalsize N\epsfxsize \picsize\fi
\epsfbox{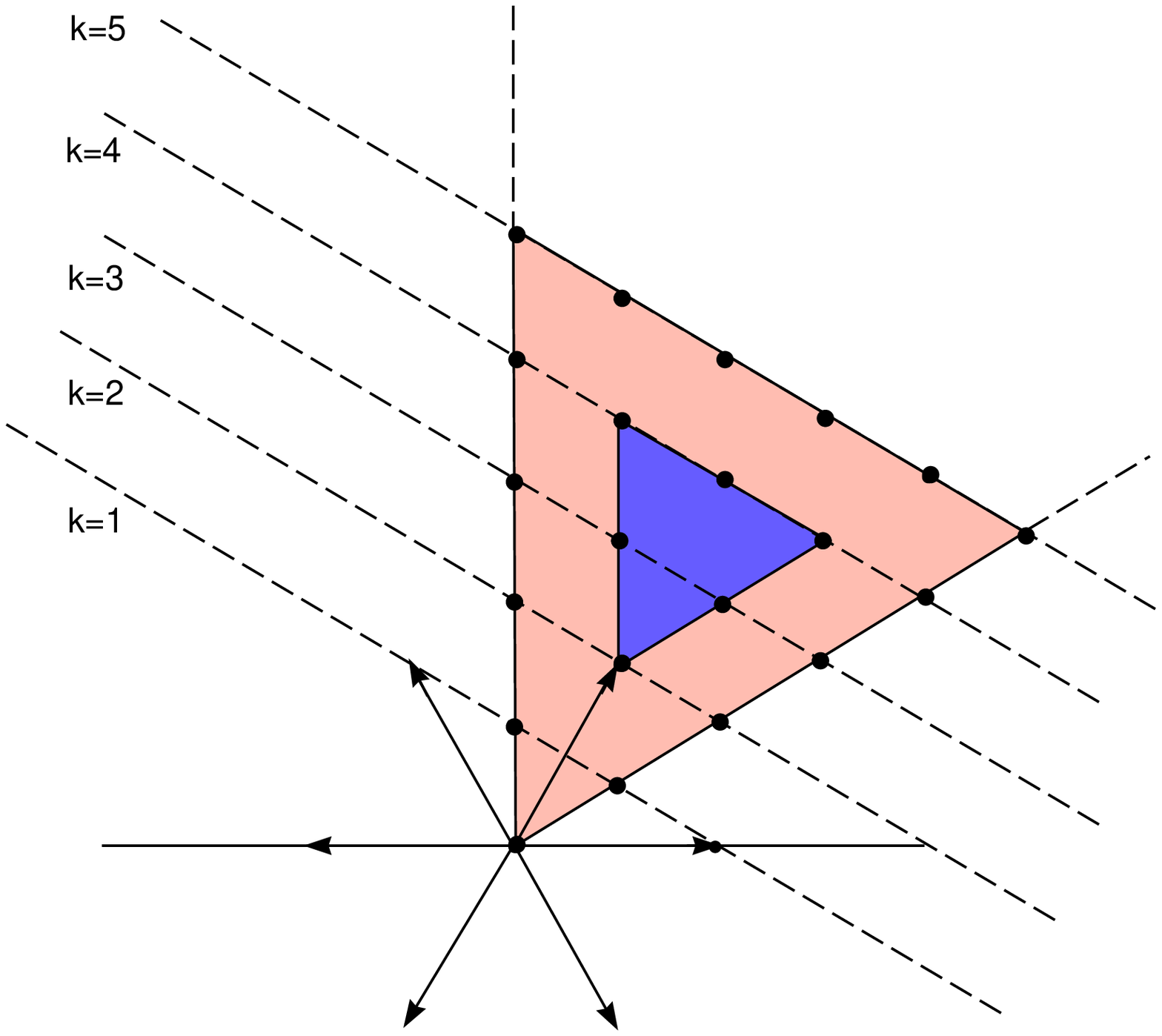}}}\fi
\vskip -2.truecm

To map the scaled, shifted weight back to the
dark triangle  we can only use affine Weyl reflections
at level $k+g$, if we want $S$ to transform in a reasonable way. These
transformations can be used to map all weights into the large triangle, but
it would seem that nothing
guarantees that we end up in the dark triangle. However, it is
not hard to see that if the scale factor $l$ is prime relative to
$M(k+g)$ (which it must be due to the Galois condition) we do indeed end up in
the dark triangle. In this way any highest weight $\mu$ is mapped to
another highest weight, possibly accompanied by a sign change of $S$.
In this way one can compute the matrices $\Pi$ defined
above. They can then be to construct elements in the integer
commutant of $S$.

In [2] we
have investigated the use of this knowledge for finding WZW modular
invariants, and --  to our surprise -- we actually found some new
invariants!

If $l$ does not satisfy the Galois condition the map $\Pi$ can not be
expected to be an automorphism. Indeed, one finds that it is
not invertible, and in many cases some or all fields are transformed outside
of the dark triangle. Those fields end up on the border of the
large triangle. We can still represent the map by a matrix $\Pi$ just as
before,
simply by ignoring the fields that are mapped to the border.
Now the matrix $\Pi$ will however be non-invertible.
 We call such
scalings {\it quasi-Galois} scalings [3].

To illustrate the difference between
Galois and quasi-Galois scalings, let us consider $G_{2,4}$. There are 9
primary fields, labelled
as follows.
$$\eqalign{0  = (0,0);  \quad       1 =(0,1)  ;
\quad      2 =  (0,2);\cr
3 = (0,3) ;   \quad
4  = (0,4) ;   \quad
5  = (1,0)  ;\cr
6  = (1,1)  ;   \quad
7 =  (1,2)   ;  \quad
8 =  (2,0)    . \cr } $$
Scaling by a factor 11, which is prime with respect to $N=3(k+g)=24$, we get
a set of signs and permutations that can be represented by the
following diagram. Here solid lines indicate a positive sign and
dashed lines a negative one.

\vskip 1.truecm

\let\picnaturalsize=N
\def\picsize{3.5in}
\def\picfilename{G2Galois.eps}
\ifx\nopictures Y\else{\ifx\epsfloaded Y\else\input epsf \fi
\let\epsfloaded=Y
\centerline{\ifx\picnaturalsize N\epsfxsize \picsize\fi
\epsfbox{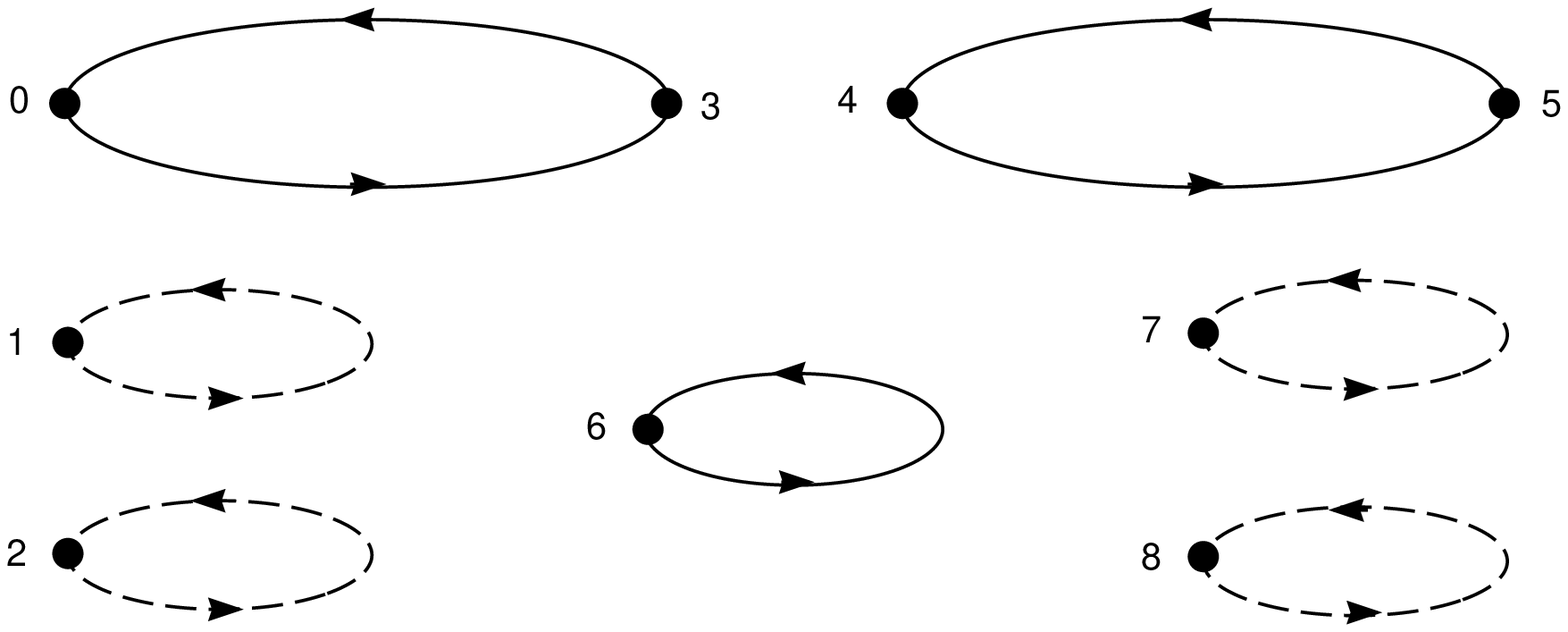}}}\fi
\vskip .7truecm
\noindent

The matrix $\Pi$ is obtained by setting to $\Pi_{ij}$ to 1(-1)
whenever there is a solid (dashed) arrow from $i$ to $j$, and to
zero if there is no line.  This matrix
is actually symmetric, and hence $\Pi$ itself is an integer
matrix commuting with $S$. It turns out that $\Pi+{\bf 1}$ is a positive
modular invariant, corresponding to the conformal embedding
$G_{2,4} \subset SO(14)$.

The next picture illustrates a quasi-Galois scaling with $l=2$, again
for $G_{2,4}$. The grey blob
represents the border of the positive Weyl chamber at level $k+g$.
All other conventions are as before. It is quite clear that
this picture represents a non-invertible map.

\let\picnaturalsize=N
\def\picsize{2.0in}
\def\picfilename{G2NonGalois.eps}
\ifx\nopictures Y\else{\ifx\epsfloaded Y\else\input epsf \fi
\let\epsfloaded=Y
\centerline{\ifx\picnaturalsize N\epsfxsize \picsize\fi
\epsfbox{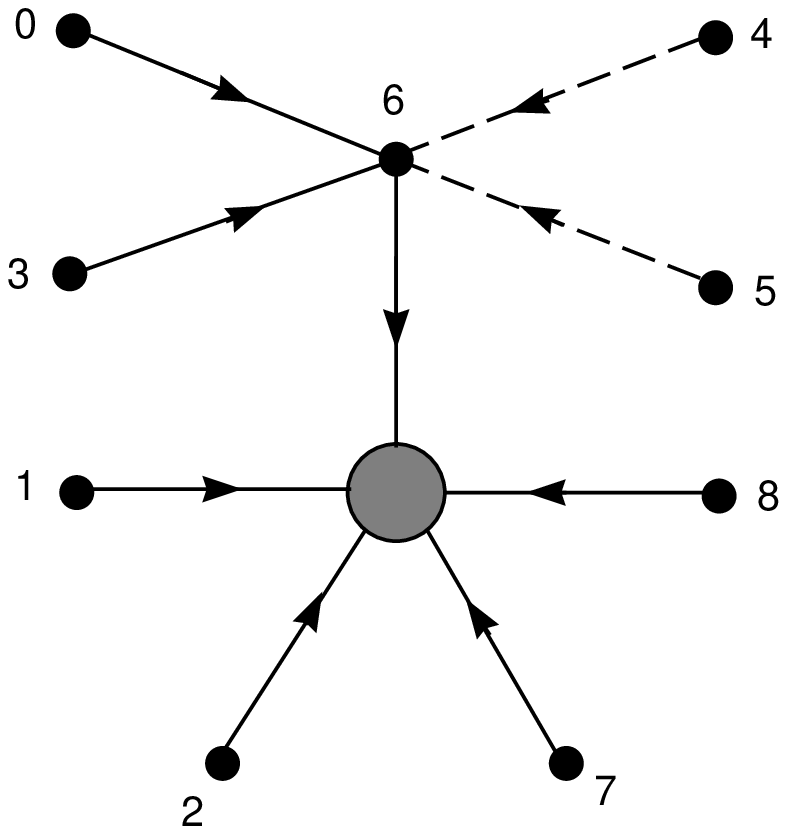}}}\fi

If we restrict ourselves to the five fields 0,3,4,5 and 6, the matrix $\Pi$
has the following form
$$\Pi = \pmatrix{ 0 & 0 & 0 & 0 & 1 \cr
0 & 0 & 0 & 0 & 1 \cr
0 & 0 & 0 & 0 & -1 \cr
0 & 0 & 0 & 0 & -1 \cr
0 & 0 & 0 & 0 & 0 \cr} \ ,
 $$
and vanishes outside this subspace.

Remarkably, even though $\Pi$ is not invertible, and though \SRel\ does not
hold, it turns out that \PiTPi\ {\it does} hold. This is non-trivial and
will be proved elsewhere [3]. An immediate consequence is that
$\Delta=\Pi + \Pi^T$ commutes with $S$. Actually, it is not this matrix
that is of most interest, but rather $\Delta^2$, which of
course also commutes with $S$. On the same subspace as before, it has
the form

$$ \pmatrix{~~1 &~~1& -1& -1 & 0 \cr
{}~~1 &~~1 & -1 & -1& 0 \cr
-1 & -1 &~~1 &~~1 & 0\cr
-1 & -1 &~~1 &~~1 &0 \cr
{}~~0 & ~~0 & ~~0 & ~~0 & 4 \cr} \ ,
 $$

This is not a good modular invariant, but it turns out that the
$G_{2,4}$ invariant described above can be obtained from it via a
GSO-like  projection. That invariant is explicitly:

$$M_{G_{2,4}} = |(0,0)+(0,3)|^2+|(0,4+1,0)|^2 + 2|(1,1)|^2 $$

The same quasi-Galois
construction turns out to work for all conformal embeddings of
a Kac-Moody algebra at level $g$ into $SO(D)$, where $D$ is the
dimension of the adjoint representation
(on the other hand, the Galois construction of the $G_{2,4}$ invariant is
special for that algebra).
In general they do not come out
directly, but one gets a non-positive $S,  T^2$ invariant from which the
form of the desired invariant can be conjectured by applying a
GSO-like projection.
This is a useful result, since it is quite hard to construct those invariants
by any other method\rlap.\foot{After [3] was first posted on ``hep-th" we
learned that Kac and Wakimoto [15] have presented -- without any proof --
formulas from which the decomposition of the identity and vector
character can be derived.  These
formulas turn out to be equivalent to our results. No examples were given,
however, and working them out with the formula in the form
given in [15] would still
be a laborious task.}
 For example, $E_{8,30}$ has 20956 primary fields,
so that computing the matrix $S$ is a horrendous task. Nevertheless we
have been able to give a very plausible conjecture for the form of the
invariant corresponding to the conformal embedding in $SO(248)$ using
a quasi-Galois scaling with $l=2$.
We will discuss the
details in [3], and we present in the appendix
the results for the
other exceptional algebras.

(Quasi)-Galois symmetry is yet another piece in the huge puzzle of classifying
and describing (rational) conformal field theories. Most problems
in this field remain
unsolved. It is likely that the full potential of (quasi)-Galois
symmetries has not been exploited yet. I hope that
they can  contribute to a more satisfactory state of affairs.

\ack

I would like to thank Christoph Schweigert and J\"urgen Fuchs for
carefully reading the manuscript.
\vfill\eject
\tenpoint
\singlespace

\appendix

$$\eqalign{M_{F_{4,9}} &= \cr
&+ |(0,0,0,0)
+ (0,0,1,6)
+ (0,0,2,1)
+ (0,1,0,0)
\cr
 &+ (0,1,1,2)
+ (0,3,0,0)
+ (1,0,0,5)
+ (1,1,0,4)|^2
\cr
{}~~~~~~~~~~~~\cr
&+ |(0,0,0,7,2)
+ (0,0,2,0,5)
+ (0,0,3,0,3)
+ (0,1,0,3,3)
\cr
 &+ (0,1,0,6,0)
+ (0,2,0,2,1)
+ (1,0,0,0,7)
+ (1,0,1,4,1)|^2
\cr
{}~~~~~~~~\cr
&+ 2 | 2 (1,1,1,1) |^2\cr}$$

$$\eqalign{M_{E_{6,12}} &= \cr
&+ |(0,0,0,0,0,0)
+ (12,0,0,0,0,0)
+ (0,0,0,0,12,0)
+ (0,0,1,0,0,0)
\cr
 &+ (9,0,1,0,0,0)
+ (0,0,1,0,9,0)
+ (0,0,2,0,3,0)
+ (3,0,2,0,0,0)
\cr
 &+ (3,0,2,0,3,0)
+ (0,1,0,0,5,2)
+ (1,2,0,1,0,0)
+ (5,0,0,2,1,1)
\cr
 &+ (0,1,0,2,1,0)
+ (5,0,0,1,0,2)
+ (1,2,0,0,5,1)
+ (0,2,0,0,1,0)
\cr
 &+ (7,0,0,2,0,0)
+ (1,0,0,0,7,2)
+ (0,2,0,0,7,0)
+ (1,0,0,2,0,0)
\cr
 &+ (7,0,0,0,1,2)
+ (1,0,3,0,1,0)
+ (1,1,1,0,3,1)
+ (1,1,1,1,1,0)
\cr
 &+ (3,0,1,1,1,1)
+ (2,0,0,1,3,1)
+ (3,1,0,0,2,1)
+ (3,1,0,1,3,0)
\cr
 &+ (2,0,1,0,2,0)
+ (5,0,1,0,2,0)
+ (2,0,1,0,5,0)
+ (4,0,0,0,4,0)
|^2 \cr
{}~~~~~~~~~~~\cr
&+ |(0,0,0,0,0,1)
+ (10,1,0,0,0,0)
+ (0,0,0,1,10,0)
+ (0,0,0,0,6,3)
\cr
 &+ (0,3,0,0,0,0)
+ (6,0,0,3,0,0)
+ (0,0,0,3,0,0)
+ (6,0,0,0,0,3)
\cr
 &+ (0,3,0,0,6,0)
+ (0,0,4,0,0,0)
+ (0,1,0,0,8,1)
+ (0,1,0,1,0,0)
\cr
 &+ (8,0,0,1,0,1)
+ (0,1,2,0,2,0)
+ (2,0,2,1,0,0)
+ (2,0,2,0,2,1)
\cr
 &+ (0,2,0,0,4,2)
+ (0,2,0,2,0,0)
+ (4,0,0,2,0,2)
+ (1,0,1,0,4,1)
\cr
 &+ (2,1,1,0,1,0)
+ (4,0,1,1,2,0)
+ (1,0,1,1,2,0)
+ (4,0,1,0,1,1)
\cr
 &+ (2,1,1,0,4,0)
+ (1,1,0,0,6,1)
+ (1,1,0,1,1,0)
+ (6,0,0,1,1,1)
\cr
 &+ (2,1,0,1,2,1)
+ (3,0,0,0,3,1)
+ (4,1,0,0,3,0)
+ (3,0,0,1,4,0)   |^2
\cr
{}~~~~~~~~~~~~~~\cr
& + 2 | 4 (1,1,1,1,1,1)   |^2 \cr } $$

$$\eqalign{
M_{E_{7,18}} &= \cr
&+ |(0,0,0,0,0,0,0)
+ (0,0,0,0,0,10,4)
+ (0,0,0,0,0,1,3)
+ (1,0,0,0,0,12,2)
\cr
 &+ (0,0,0,0,1,16,0)
+ (0,0,0,0,4,0,2)
+ (4,0,0,0,0,7,1)
+ (0,0,0,0,5,0,0)
\cr
 &+ (0,0,0,1,0,0,1)
+ (0,0,0,1,0,0,3)
+ (0,1,0,0,0,10,2)
+ (0,1,0,0,0,0,0)
\cr
 &+ (0,0,0,1,1,0,1)
+ (0,2,0,0,0,12,0)
+ (0,0,0,3,0,4,1)
+ (0,3,0,0,0,4,0)
\cr
 &+ (0,0,0,6,0,0,0)
+ (0,0,1,0,0,14,0)
+ (1,0,1,0,0,8,2)
+ (0,0,1,0,1,1,1)
\cr
 &+ (2,0,1,0,0,10,0)
+ (3,0,1,0,0,5,1)
+ (0,0,1,0,3,2,0)
+ (0,3,1,0,0,2,0)
\cr
 &+ (0,0,2,0,0,1,1)
+ (0,0,2,0,0,8,0)
+ (0,0,2,0,3,0,0)
+ (0,0,3,0,0,6,0)
\cr
 &+ (0,1,0,0,2,0,2)
+ (2,0,0,1,0,8,1)
+ (0,1,0,0,3,0,0)
+ (0,1,0,1,2,2,1)
\cr
 &+ (2,1,0,1,0,5,0)
+ (0,2,0,1,0,2,1)
+ (0,1,0,2,0,6,0)
+ (0,1,0,4,0,2,0)
\cr
 &+ (1,0,1,1,0,6,1)
+ (0,1,1,0,1,2,0)
+ (2,1,1,1,0,3,0)
+ (0,1,1,2,0,4,0)
\cr
 &+ (0,1,2,0,1,0,0)
+ (2,0,0,2,0,3,1)
+ (0,2,0,0,2,4,0)
+ (0,2,0,2,2,0,0)
\cr
 &+ (0,3,0,2,0,0,0)
+ (0,2,1,0,2,2,0)
+ (2,1,0,3,0,1,0)
+ (0,4,0,0,3,0,0)
\cr
 &+ (0,5,0,0,1,0,0)
+ (3,0,0,0,1,6,2)
+ (1,0,0,0,3,1,1)
+ (4,0,0,0,1,8,0)
\cr
 &+ (1,2,0,0,1,3,1)
+ (1,0,0,2,1,4,0)
+ (1,0,1,0,2,1,1)
+ (2,0,1,0,1,6,0)
\cr
 &+ (1,0,1,2,1,2,0)
+ (2,0,2,0,1,4,0)
+ (1,1,0,1,1,4,1)
+ (1,1,0,1,1,3,0)
\cr
 &+ (1,3,0,1,1,1,0)
+ (1,1,1,1,1,1,0)
+ (1,2,0,2,1,2,0)
+ (2,0,0,4,1,0,0) |^2 \cr
&~~~~\cr
&+ |(0,0,0,0,0,18,0)
+ (0,0,0,0,0,0,4)
+ (0,0,0,0,0,11,3)
+ (0,0,0,0,1,0,2)
\cr
 &+ (1,0,0,0,0,0,0)
+ (4,0,0,0,0,6,2)
+ (0,0,0,0,4,1,1)
+ (5,0,0,0,0,8,0)
\cr
 &+ (0,1,0,0,0,13,1)
+ (0,1,0,0,0,9,3)
+ (0,0,0,1,0,1,2)
+ (0,0,0,1,0,15,0)
\cr
 &+ (1,1,0,0,0,11,1)
+ (0,0,0,2,0,0,0)
+ (0,3,0,0,0,3,1)
+ (0,0,0,3,0,5,0)
\cr
 &+ (0,6,0,0,0,0,0)
+ (0,0,1,0,0,0,0)
+ (0,0,1,0,1,0,2)
+ (1,0,1,0,0,9,1)
\cr
 &+ (0,0,1,0,2,0,0)
+ (0,0,1,0,3,1,1)
+ (3,0,1,0,0,6,0)
+ (0,0,1,3,0,3,0)
\cr
 &+ (0,0,2,0,0,7,1)
+ (0,0,2,0,0,2,0)
+ (3,0,2,0,0,4,0)
+ (0,0,3,0,0,0,0)
\cr
 &+ (2,0,0,1,0,7,2)
+ (0,1,0,0,2,1,1)
+ (3,0,0,1,0,9,0)
+ (2,1,0,1,0,4,1)
\cr
 &+ (0,1,0,1,2,3,0)
+ (0,1,0,2,0,5,1)
+ (0,2,0,1,0,3,0)
+ (0,4,0,1,0,1,0)
\cr
 &+ (0,1,1,0,1,1,1)
+ (1,0,1,1,0,7,0)
+ (0,1,1,1,2,1,0)
+ (0,2,1,1,0,1,0)
\cr
 &+ (1,0,2,1,0,5,0)
+ (0,2,0,0,2,3,1)
+ (2,0,0,2,0,4,0)
+ (2,2,0,2,0,2,0)
\cr
 &+ (0,2,0,3,0,3,0)
+ (2,0,1,2,0,2,0)
+ (0,3,0,1,2,1,0)
+ (3,0,0,4,0,0,0)
\cr
 &+ (1,0,0,5,0,1,0)
+ (1,0,0,0,3,0,2)
+ (3,0,0,0,1,7,1)
+ (1,0,0,0,4,0,0)
\cr
 &+ (1,0,0,2,1,3,1)
+ (1,2,0,0,1,4,0)
+ (2,0,1,0,1,5,1)
+ (1,0,1,0,2,2,0)
\cr
 &+ (1,2,1,0,1,2,0)
+ (1,0,2,0,2,0,0)
+ (1,1,0,1,1,2,1)
+ (1,1,0,1,1,5,0)
\cr
 &+ (1,1,0,3,1,1,0)
+ (1,1,1,1,1,3,0)
+ (1,2,0,2,1,0,0)
+ (1,4,0,0,2,0,0) |^2 \cr
{}~~~~~~~~~~~\cr
&+ | 8 (1,1,1,1,1,1,1)|^2}
$$

$$\eqalign{ M_{E_{8,30}} =  \cr
 &~~~|(0,0,0,0,0,0,0,0)
+ (0,0,0,0,0,0,0,3)
+ (0,0,0,0,0,0,0,10)
+ (0,0,0,0,0,0,7,0)
\cr
 &+ (0,0,0,0,0,1,0,1)
+ (0,0,0,0,0,1,1,1)
+ (0,0,0,0,1,0,1,1)
+ (0,0,0,0,1,0,5,0)
\cr
 &+ (0,0,0,0,4,0,0,0)
+ (0,0,0,1,0,0,0,0)
+ (0,0,0,1,0,0,3,0)
+ (0,0,0,1,0,4,0,0)
\cr
 &+ (0,0,0,1,1,0,1,0)
+ (0,0,0,2,0,0,4,0)
+ (0,0,0,3,0,0,0,0)
+ (0,0,0,3,0,0,0,4)
\cr
 &+ (0,0,0,6,0,0,0,0)
+ (0,0,1,0,0,0,3,1)
+ (0,0,1,0,0,2,3,0)
+ (0,0,1,1,0,1,1,0)
\cr
 &+ (0,0,2,0,0,3,0,1)
+ (0,0,2,0,1,0,3,0)
+ (0,0,2,0,3,0,0,0)
+ (0,0,2,1,0,1,0,0)
\cr
 &+ (0,0,3,0,0,0,0,6)
+ (0,0,4,0,0,0,0,1)
+ (0,0,4,0,0,0,3,0)
+ (0,1,0,0,0,0,0,0)
\cr
 &+ (0,1,0,0,0,0,5,0)
+ (0,1,0,0,1,0,3,0)
+ (0,1,0,0,2,2,0,0)
+ (0,1,0,1,0,2,2,0)
\cr
 &+ (0,1,0,1,2,0,0,2)
+ (0,1,0,2,0,0,2,0)
+ (0,1,0,4,0,0,0,2)
+ (0,1,1,0,0,2,1,0)
\cr
 &+ (0,1,1,2,0,0,0,4)
+ (0,1,2,0,0,1,2,1)
+ (0,1,2,0,1,0,1,0)
+ (0,1,2,0,1,2,0,0)
\cr
 &+ (0,1,4,0,0,0,1,0)
+ (0,2,0,0,2,0,2,0)
+ (0,2,0,1,0,2,0,0)
+ (0,2,0,1,0,2,0,2)
\cr
 &+ (0,2,0,2,2,0,0,0)
+ (0,2,1,0,2,0,0,2)
+ (0,2,2,0,0,1,0,1)
+ (0,2,2,0,1,0,2,0)
\cr
 &+ (0,3,0,0,2,0,0,0)
+ (0,3,0,1,0,0,2,2)
+ (0,3,0,2,0,2,0,0)
+ (0,3,1,0,0,2,0,2)
\cr
 &+ (0,3,2,0,1,0,0,0)
+ (0,4,0,0,3,0,0,0)
+ (0,4,0,1,0,0,0,2)
+ (0,4,0,2,0,0,2,0)
\cr
 &+ (0,4,1,0,0,0,2,2)
+ (0,5,0,0,1,2,0,0)
+ (0,5,0,2,0,0,0,0)
+ (0,5,1,0,0,0,0,2)
\cr
 &+ (0,6,0,0,1,0,2,0)
+ (0,7,0,0,1,0,0,0)
+ (0,8,0,0,0,0,3,0)
+ (0,9,0,0,0,0,1,0)
\cr
 &+ (1,0,0,0,0,0,5,1)
+ (1,0,0,0,0,4,1,0)
+ (1,0,0,1,0,1,3,0)
+ (1,0,1,0,0,0,0,8)
\cr
 &+ (1,0,1,0,1,2,1,0)
+ (1,0,1,0,2,1,0,1)
+ (1,0,1,1,0,0,0,6)
+ (1,0,1,1,0,1,2,0)
\cr
 &+ (1,0,1,2,1,0,0,2)
+ (1,0,3,0,0,0,2,1)
+ (1,0,3,0,0,2,1,0)
+ (1,1,0,0,0,3,1,0)
\cr
 &+ (1,1,0,1,1,0,0,4)
+ (1,1,1,0,1,1,1,0)
+ (1,1,1,0,1,1,1,1)
+ (1,1,1,1,1,1,0,1)
\cr
 &+ (1,1,3,0,0,1,1,0)
+ (1,2,0,0,1,1,0,3)
+ (1,2,0,2,1,0,0,2)
+ (1,2,1,0,1,0,1,1)
\cr
 &+ (1,2,1,1,0,1,1,1)
+ (1,3,0,0,0,1,1,3)
+ (1,3,0,1,1,1,0,1)
+ (1,3,1,1,0,0,1,1)
\cr
 &+ (1,4,0,0,0,0,1,3)
+ (1,4,0,1,0,1,1,1)
+ (1,5,0,1,0,0,1,1)
+ (1,6,0,0,0,2,1,0)
\cr
 &+ (1,7,0,0,0,1,1,0)
+ (2,0,0,0,3,0,1,0)
+ (2,0,0,1,0,3,0,0)
+ (2,0,0,2,0,1,0,3)
\cr
 &+ (2,0,0,4,1,0,0,0)
+ (2,0,2,0,0,2,0,1)
+ (2,0,2,0,1,0,0,4)
+ (2,0,2,0,2,0,1,0)
\cr
 &+ (2,1,0,0,2,1,0,0)
+ (2,1,0,1,1,0,1,2)
+ (2,1,0,3,0,1,0,1)
+ (2,1,1,1,0,1,0,3)
\cr
 &+ (2,1,2,0,1,1,0,0)
+ (2,2,0,1,0,1,0,2)
+ (2,2,0,2,1,0,1,0)
+ (2,2,1,0,1,0,1,2)
\cr
 &+ (2,3,0,2,0,1,0,0)
+ (2,3,1,0,0,1,0,2)
+ (2,4,0,0,2,0,1,0)
+ (2,5,0,0,1,1,0,0)
\cr
 &+ (3,0,0,0,1,0,0,6)
+ (3,0,1,0,0,1,0,5)
+ (3,0,1,0,2,0,0,1)
+ (3,0,1,2,0,0,1,2)
\cr
 &+ (3,1,0,1,0,0,1,4)
+ (3,1,1,1,1,0,0,1)
+ (3,2,0,0,1,0,0,3)
+ (3,2,0,2,0,0,1,2)
\cr
 &+ (3,3,0,1,1,0,0,1)
+ (4,0,0,2,0,0,0,3)
+ (4,0,0,4,0,0,1,0)
+ (4,0,2,0,0,0,1,4)
\cr
 &+ (4,1,0,3,0,0,0,1)
+ (4,1,1,1,0,0,0,3)
+ (5,0,0,0,0,0,1,6)
+ (5,0,1,0,0,0,0,5) |^2\cr
&+ \ldots \cr}
$$
$$ \eqalign{
\ldots&+|(0,0,0,0,0,0,1,2)
+ (0,0,0,0,0,0,6,1)
+ (0,0,0,0,0,1,0,2)
+ (0,0,0,0,0,2,0,0)
\cr
 &+ (0,0,0,0,0,5,0,0)
+ (0,0,0,0,1,0,0,0)
+ (0,0,0,0,1,0,2,0)
+ (0,0,0,0,2,0,0,0)
\cr
 &+ (0,0,0,1,0,0,2,1)
+ (0,0,0,1,0,1,4,0)
+ (0,0,0,2,0,1,0,0)
+ (0,0,1,0,0,0,0,0)
\cr
 &+ (0,0,1,0,0,0,4,0)
+ (0,0,1,0,1,0,2,0)
+ (0,0,1,0,1,3,0,0)
+ (0,0,1,0,3,0,0,1)
\cr
 &+ (0,0,1,1,0,1,3,0)
+ (0,0,1,2,0,0,1,0)
+ (0,0,1,3,0,0,0,3)
+ (0,0,2,0,0,0,0,7)
\cr
 &+ (0,0,2,0,0,2,0,0)
+ (0,0,3,0,0,0,3,1)
+ (0,0,3,0,0,3,0,0)
+ (0,0,3,0,1,0,0,0)
\cr
 &+ (0,0,5,0,0,0,0,0)
+ (0,1,0,0,0,0,0,9)
+ (0,1,0,0,0,0,4,1)
+ (0,1,0,0,0,3,2,0)
\cr
 &+ (0,1,0,1,0,1,2,0)
+ (0,1,0,2,0,0,0,5)
+ (0,1,1,0,1,1,2,0)
+ (0,1,1,0,1,2,0,1)
\cr
 &+ (0,1,1,1,0,1,1,0)
+ (0,1,1,1,2,0,0,1)
+ (0,1,3,0,0,0,1,1)
+ (0,1,3,0,0,1,2,0)
\cr
 &+ (0,2,0,0,0,3,0,0)
+ (0,2,0,0,2,0,0,3)
+ (0,2,0,3,0,0,0,3)
+ (0,2,1,0,1,0,2,1)
\cr
 &+ (0,2,1,0,1,1,0,0)
+ (0,2,1,1,0,2,0,1)
+ (0,2,3,0,0,1,0,0)
+ (0,3,0,0,0,2,0,3)
\cr
 &+ (0,3,0,1,2,0,0,1)
+ (0,3,1,0,1,0,0,1)
+ (0,3,1,1,0,0,2,1)
+ (0,4,0,0,0,0,2,3)
\cr
 &+ (0,4,0,1,0,2,0,1)
+ (0,4,1,1,0,0,0,1)
+ (0,5,0,0,0,0,0,3)
+ (0,5,0,1,0,0,2,1)
\cr
 &+ (0,6,0,0,0,3,0,0)
+ (0,6,0,1,0,0,0,1)
+ (0,7,0,0,0,1,2,0)
+ (0,8,0,0,0,1,0,0)
\cr
 &+ (0,10,0,0,0,0,0,0)
+ (1,0,0,0,0,0,0,0)
+ (1,0,0,0,0,0,6,0)
+ (1,0,0,0,1,0,4,0)
\cr
 &+ (1,0,0,0,3,1,0,0)
+ (1,0,0,1,0,3,1,0)
+ (1,0,0,2,0,0,3,0)
+ (1,0,0,2,1,0,0,3)
\cr
 &+ (1,0,0,5,0,0,0,1)
+ (1,0,1,0,0,2,2,0)
+ (1,0,2,0,0,2,1,1)
+ (1,0,2,0,1,0,2,0)
\cr
 &+ (1,0,2,0,2,1,0,0)
+ (1,0,2,1,0,0,0,5)
+ (1,0,4,0,0,0,2,0)
+ (1,1,0,0,2,1,1,0)
\cr
 &+ (1,1,0,1,0,2,1,0)
+ (1,1,0,1,1,1,0,2)
+ (1,1,0,3,1,0,0,1)
+ (1,1,1,1,1,0,0,3)
\cr
 &+ (1,1,2,0,0,1,1,1)
+ (1,1,2,0,1,1,1,0)
+ (1,2,0,0,2,0,1,0)
+ (1,2,0,1,0,1,1,2)
\cr
 &+ (1,2,0,2,1,1,0,0)
+ (1,2,1,0,1,1,0,2)
+ (1,2,2,0,1,0,1,0)
+ (1,3,0,1,0,0,1,2)
\cr
 &+ (1,3,0,2,0,1,1,0)
+ (1,3,1,0,0,1,1,2)
+ (1,4,0,0,2,1,0,0)
+ (1,4,0,2,0,0,1,0)
\cr
 &+ (1,4,1,0,0,0,1,2)
+ (1,5,0,0,1,1,1,0)
+ (1,6,0,0,1,0,1,0)
+ (1,8,0,0,0,0,2,0)
\cr
 &+ (2,0,0,0,0,4,0,0)
+ (2,0,0,1,0,0,0,7)
+ (2,0,1,0,1,0,0,5)
+ (2,0,1,0,1,2,0,0)
\cr
 &+ (2,0,1,0,2,0,1,1)
+ (2,0,1,2,0,1,0,2)
+ (2,0,3,0,0,2,0,0)
+ (2,1,0,1,0,1,0,4)
\cr
 &+ (2,1,1,0,1,1,0,1)
+ (2,1,1,1,1,0,1,1)
+ (2,2,0,0,1,0,1,3)
+ (2,2,0,2,0,1,0,2)
\cr
 &+ (2,2,1,1,0,1,0,1)
+ (2,3,0,0,0,1,0,3)
+ (2,3,0,1,1,0,1,1)
+ (2,4,0,1,0,1,0,1)
\cr
 &+ (2,6,0,0,0,2,0,0)
+ (3,0,0,0,3,0,0,0)
+ (3,0,0,2,0,0,1,3)
+ (3,0,0,4,0,1,0,0)
\cr
 &+ (3,0,2,0,0,1,0,4)
+ (3,0,2,0,2,0,0,0)
+ (3,1,0,1,1,0,0,2)
+ (3,1,0,3,0,0,1,1)
\cr
 &+ (3,1,1,1,0,0,1,3)
+ (3,2,0,2,1,0,0,0)
+ (3,2,1,0,1,0,0,2)
+ (3,4,0,0,2,0,0,0)
\cr
 &+ (4,0,0,0,0,1,0,6)
+ (4,0,1,0,0,0,1,5)
+ (4,0,1,2,0,0,0,2)
+ (4,1,0,1,0,0,0,4)
\cr
 &+ (4,2,0,2,0,0,0,2)
+ (5,0,0,4,0,0,0,0)
+ (5,0,2,0,0,0,0,4)
+ (6,0,0,0,0,0,0,6)|^2
\cr
{}~~~~~~~~~~~~~\cr
&+2| 8(1,1,1,1,1,1,1,1)|^2 \cr }$$
\eject
\twelvepoint
\normalspace
\centerline{\fourteenrm REFERENCES}
\vskip 1.truecm
\refitem {[{1}] }
\obeyendofline J. Fuchs, B. Gato-Rivera, A.N. Schellekens and C. Schweigert,
\plt B334 (1994) 113.
\ignoreendofline
\refitem {[{2}] }
\obeyendofline  J. Fuchs, A.N. Schellekens and C. Schweigert, preprint
NIKHEF-94/31.
\ignoreendofline
\refitem {[{3}] }
\obeyendofline J. Fuchs, A.N. Schellekens and C. Schweigert,  preprint
NIKHEF-94/37.
\ignoreendofline
\refitem {[{4}] }
\obeyendofline  A.~Cappelli, C.~Itzykson and J.-B.~Zuber, \nup 280 (1987)
445;\rB \cmp 113 (1987) 1.
\ignoreendofline
\refitem {[{5}] }
\obeyendofline T. Gannon, \cmp {161}(1994){233}.
\ignoreendofline
\refitem {[{6}] }
\obeyendofline  A.N.~Schellekens and S.~Yankielowicz, \nup 327 (1989) 673; \rB
\plt B227 (1989) 387.
\ignoreendofline
\refitem {[{7}] }
\obeyendofline D.~Bernard, \nup 288 (1987) 628; D.~Altschuler, J.~Lacki and
P.~Zaugg, \rB \plt B205 (1988) 281; G.~Felder, K.~Gawedzki and A.~Kupiainen,
\cmp 117 (1988) 127.
\ignoreendofline
\refitem {[{8}] }
\obeyendofline J.~Fuchs, \cmp 136 (1991) 345.
\ignoreendofline
\refitem {[{9}] }
\obeyendofline E.~Verlinde, \nup 300 (1988) 360.
\ignoreendofline
\refitem {[{10}] }
\obeyendofline A.N.~Schellekens and S.~Yankielowicz, Int.~J.~Mod.~ Phys.~\und
{A5} (1990) 2903.
\ignoreendofline
\refitem {[{11}] }
\obeyendofline J.~Fuchs and C.~Schweigert, Ann. Phys. 234 (1994) 102.
\ignoreendofline
\refitem {[{12}] }
\obeyendofline W.~Eholzer, Int. J. Mod. Phys. 8 (1993) {3495}.
\ignoreendofline
\refitem {[{13}] }
\obeyendofline J. de Boer and J. Goeree, \cmp 139 (1991) 267.
\ignoreendofline
\refitem {[{14}] }
\obeyendofline A. Coste and T. Gannon, \plt B323 (1994) 316.
\ignoreendofline
\refitem {[{15}] }
\obeyendofline V.Kac and M. Wakimoto, Adv. Math. 70 (1988) 156.
\ignoreendofline
\end